\documentclass[prc,preprint,showpacs,preprintnumbers,amsmath,amssymb]{revtex4}

\usepackage[mathscr]{eucal}
\usepackage{graphicx}
\def\slr#1{\setbox0=\hbox{$#1$}           
   \dimen0=\wd0                                 
   \setbox1=\hbox{/} \dimen1=\wd1               
   \ifdim\dimen0>\dimen1                        
      \rlap{\hbox to \dimen0{\hfil/\hfil}}      
      #1                                        
   \else                                        
      \rlap{\hbox to \dimen1{\hfil$#1$\hfil}}   
      /                                         
   \fi}

\def\gev#1{ GeV${}^{#1}$}
\def\be{\begin{eqnarray}}
\def\ee{\end{eqnarray}}

\renewcommand{\theequation}%
    {\arabic{section}.\arabic{equation}}
\makeatletter \@addtoreset{equation}{section} \makeatother

\begin{document}

\preprint{BCCNT: 02/111/317}

\title{Calculation of the Excitations of Dense Quark Matter at Zero Temperature}

\author{Bing He}
\author{Hu Li}
\author{C. M. Shakin}
 \email[email:]{casbc@cunyvm.cuny.edu}
\author{Qing Sun}

\affiliation{%
Department of Physics and Center for Nuclear Theory\\
Brooklyn College of the City University of New York\\
Brooklyn, New York 11210
}%

\date{November, 2002}

\begin{abstract}
Recently there has been a great deal of interest in studying the
properties of dense quark matter, with particular reference to
diquark condensates and color superconductivity. In the present
work we report calculations made for the excitations of quark
matter for relatively low densities of the deconfined phase and in
the absence of meson or diquark condensation. Here, we are
interested in elucidating the role of ``Pauli blocking", as such
blocking affects the calculation vector, scalar and pseudoscalar
$q\bar q$ excitations. As a byproduct of our analysis, we extend
our calculations to higher densities and explore some consequences
of the use of density-dependent coupling parameters for the
Nambu--Jona-Lasinio model. (Such density-dependent parameters have
been used in some of our previous work.) For our analysis made at
large values of the matter density, we assume that at about 13
times nuclear matter density quark matter has only minimal
nonperturbative interactions. At that high density we compare the
result for hadronic current correlation functions calculated with
density-dependent and density-independent NJL coupling constants.
We find evidence for the use of density-dependent parameters,
since the results with the density-independent constants do not go
over to the perturbative description which we assume to be correct
for $\rho\simeq 6\rho_c$, where $\rho_c$ is the matter density for
the finite-density confinement-deconfinement transition. The use
of density-dependent coupling constants in the study of diquark
condensates and color superconductivity has not been explored as
yet, and is a topic requiring further investigation, particularly
given the strong interest in the properties of dense quark matter
and color superconductivity.
\end{abstract}


\maketitle

\section{INTRODUCTION}

At present there is a great deal of interest in exploring the
properties of quantum chromodynamics (QCD) at high temperature and
baryon density. We may quote Renk, Schneider and Weise
[1]:``Evidently, investigating the changes of spectral
distributions of pseudoscalar and vector (as well as axial-vector)
excitations of the QCD vacuum with changing temperatures and
baryon densities, from moderate to extreme, is a key to
understanding QCD thermodynamics, its phases and symmetry breaking
patterns." Studies related to that program may carried forward
using lattice simulations of QCD or by using effective
Lagrangians.

A good deal is known concerning the properties of QCD at finite
temperature. However, the properties of QCD at finite matter
density are not well known, since the lattice simulations of QCD
at finite chemical potential are only in a relatively early state
of development [2-7]. In earlier works we have used a generalized
Nambu--Jona-Lasinio (NJL) model to study the
confinement-deconfinement transition for light mesons at both
finite temperature [8] and finite density [9]. We have also
studied the excitations of the quark-gluon plasma by calculating
hadronic current correlation functions at finite temperature [10,
11]. In the present work we report upon calculations of such
correlators at finite matter density in the deconfined phase. As
in the large number of applications of the NJL model and related
models to the study of high-density matter [12-15], we do not
consider explicit gluon degrees of freedom. Our calculations do
not take into account the formation of diquark condensates
associated with color superconductivity. One reason for not
studying diquark condensates and related matters at this time is
that we have some concerns as to the use of the NJL model with
constant values of the coupling constants at high density or high
temperature. It is easier to discuss the matter of
temperature-dependent effective coupling constants [8, 10] than to
discuss density-dependent coupling constants [9], since a good
deal is known about QCD thermodynamics at finite temperature. We
denote the critical temperature for the confinement-deconfinement
transition as $T_c$ and have used $T_c=170$ MeV in our work [8,
10]. From studies of a pure gluon system in QCD, it is found that
at high temperatures one approaches a weakly interacting system
slowly, with some nonperturbative effects still present at
temperatures $T=3T_c$ or 4$T_c$ [16]. For definiteness, we assume
that we can neglect nonperturbative effects for $T\gtrsim6T_c$.
For these high temperature we would expect to be able to calculate
hadronic current correlators using the NJL model and obtain the
result expected in lowest-order perturbation theory at energies
for which the effective Lagrangian may be used. (In that regard,
we limit our applications to $P^2<4$\gev2.) However, we found in
our studies that the calculated hadronic current correlations
functions had resonant features for $T\sim6T_c$, when
\emph{temperature-independent} coupling constants were used [11].
That was one of a number of reasons that we replaced the
temperature-independent coupling constants of the NJL model by
$G(T)=G[1-0.17T/T_c]$. (We remark that other dependence on the
parameter $T/T_c$ may be assumed, however, we have only explored
the linear dependence described here.) By analogy, we may
introduce $G(\rho)=G[1-\beta\rho/\rho_c]$ where $\rho_c$ is the
critical density for the confinement-deconfinement transition. In
our earlier work [9] we took $\rho_c=2.25\rho_{NM}$, where
$\rho_{NM}$ is the density of nuclear matter. Our introduction of
the parameter $\beta$ was not done in a systematic fashion in Ref.
[9], but the values used were $\beta\simeq0.18$. For the present
work, we have used the same value of $\rho_c=2.25\rho_{NM}$ and
have taken $\beta=0.17$. Therefore, for $\rho\gtrsim6\rho_c$, we
have assumed that the system has limited nonperturbative features.
Since we are not describing diquark condensation in this work, we
might limit ourselves to consideration of $\rho$ values larger
than, but not too different from $\rho_c$. However, here we will
also study the larger values of $\rho\lesssim6\rho_c$ in order to
gain information about a possible density-dependence of the NJL
coupling constants, in analogy to what was done in the case of our
finite temperature studies [8, 10, 11].

Our calculations are made for zero values of the chemical
potential. Thus, we can present our results for hadronic current
correlators for various values of $\rho/\rho_c$. When calculations
are made in that manner, we can present a particularly transparent
discussion of the role of ``Pauli blocking".

Our model for quark matter is that of two ideal Fermi gases of up
and down quarks with Fermi momentum $p_F$. The density of the
quark matter is given by $\rho_q=(2N_c/3\pi^2)p_F^3$, where
$N_c=3$ is the number of colors. That expression may be put in
contrast to the expression for the density of nuclear matter,
$\rho_{NM}=(2/3\pi^2)k_F^3$, where $k_F$ is the Fermi momentum of
the ideal gas of nucleons. With our form of $G(\rho)$ we see that,
in our model, we have assumed that at
$\rho_q=5.88\rho_c=13.2\rho_{NM}$ the system is weakly
interacting, since $G(\rho)=0$ at that density. We may ask if we
can apply the NJL model at such high densities. When
$\rho_q=13.2\rho_{NM}$, we find $p_F=1.64k_F=0.44$ GeV. Here we
have used $k_F=0.268$ GeV for nuclear matter. Since a typical
(sharp) three-momentum cutoff for the standard NJL model is 0.631
GeV [11, 17], we see that it is still possible to use the NJL
model at the largest density considered here. (It is possible to
extend the range of application by using a larger momentum cutoff
for the NJL model. However, the cutoff is usually fixed by fitting
the pion decay constant, so that the range of variation of the
cutoff parameter is limited [17-19].)

In this work we study the correlators which involve the excitation
of states with the quantum numbers of the $\rho$, $\pi$, $f_0$ and
$\eta$ mesons. The organization of our work is as follows. In Sec.
II we provide expressions for the hadronic current correlators for
the four cases considered and stress the role of Pauli blocking in
modifying these expressions. In Sec. III we provide results for
the imaginary parts of the correlators $C_\pi(P^2)$ and
$C_\rho(P^2)$ for several values of $\rho/\rho_c$. In Sec. IV we
present the values obtained for the correlators of flavor octet
currents, $C_{\eta88}(P^2)$ and $C_{f_088}(P^2)$. In Sec. V we
compare the results for the correlators at $\rho/\rho_c=5.88$
obtained with constant NJL coupling constants with those obtained
for the density-dependent coupling constants used in this work and
in our earlier work [9]. Finally, in Sec. VI we present some
further discussions and conclusions. In the following discussion
$\rho$ will represent the density of quark matter. However, we
will sometimes use the notation $\rho_q$ for that quantity.

\section{calculation of hadronic current correlation functions}

For ease of reference, we present the Lagrangian of our
generalized NJL model that incorporates a covariant model of
confinement [20-24]\be {\cal L}=&&\bar q(i\slr
\partial-m^0)q +\frac{G_S}{2}\sum_{i=0}^8[
(\bar q\lambda^iq)^2+(\bar qi\gamma_5 \lambda^iq)^2]\nonumber\\
&&-\frac{G_V}{2}\sum_{i=0}^8[
(\bar q\lambda^i\gamma_\mu q)^2+(\bar q\lambda^i\gamma_5 \gamma_\mu q)^2]\nonumber\\
&& +\frac{G_D}{2}\{\det[\bar q(1+\gamma_5)q]+\det[\bar
q(1-\gamma_5)q]\} \nonumber\\
&&+ {\cal L}_{conf}\,. \ee Here the $\lambda^i(i=0,\cdots, 8)$ are
the Gell-Mann matrices, with $\lambda^0=\sqrt{2/3}\mathbf{\,1}$,
$m^0=\mbox{diag}\,(m_u^0, m_d^0, m_s^0)$ is a matrix of current
quark masses and ${\cal L}_{conf}$ denotes our model of
confinement. The fourth term on the right is the 't Hooft
interaction. (The effective coupling constants in the channels
with $\pi$, $f_0$ or $\eta$ quantum numbers are linear
combinations of $G_S$ and $G_D$ [17].)

In this work we make use of the density-dependent constituent
quark masses that were calculated in Refs. [9, 25] using a
mean-field approximation. The up (or down) quark mass $m_u(\rho)$
is rather small for $\rho>\rho_c$ and has relatively little effect
when calculating the hadronic current correlators. The strange
quark mass for $\rho\gtrsim\rho_c$ is approximately constant with
$m_s(\rho)\simeq440$ MeV [9]. (The difference between the behavior
of the up and down quark mass and that of the strange quark mass
is due to the fact that the quark matter we consider is
nonstrange.)

In Fig. 1a we show the basic vacuum-polarization diagram of the
NJL model, where the lines denote either constituent quarks or
antiquarks. (In Fig. 1b we show the introduction of a confinement
vertex which is needed in our studies for $T<T_c$ or
$\rho<\rho_c$. Since we are considering the deconfined phase, we
can disregard the confinement vertex for this work.) With
reference to Fig. 1, we denote the quark momentum as $P/2+k$ and
the antiquark momentum as $-P/2+k$, and work in the frame where
$\vec P=0$.

\begin{figure}
\includegraphics[bb=0 0 600 300, angle=0, scale=0.4]{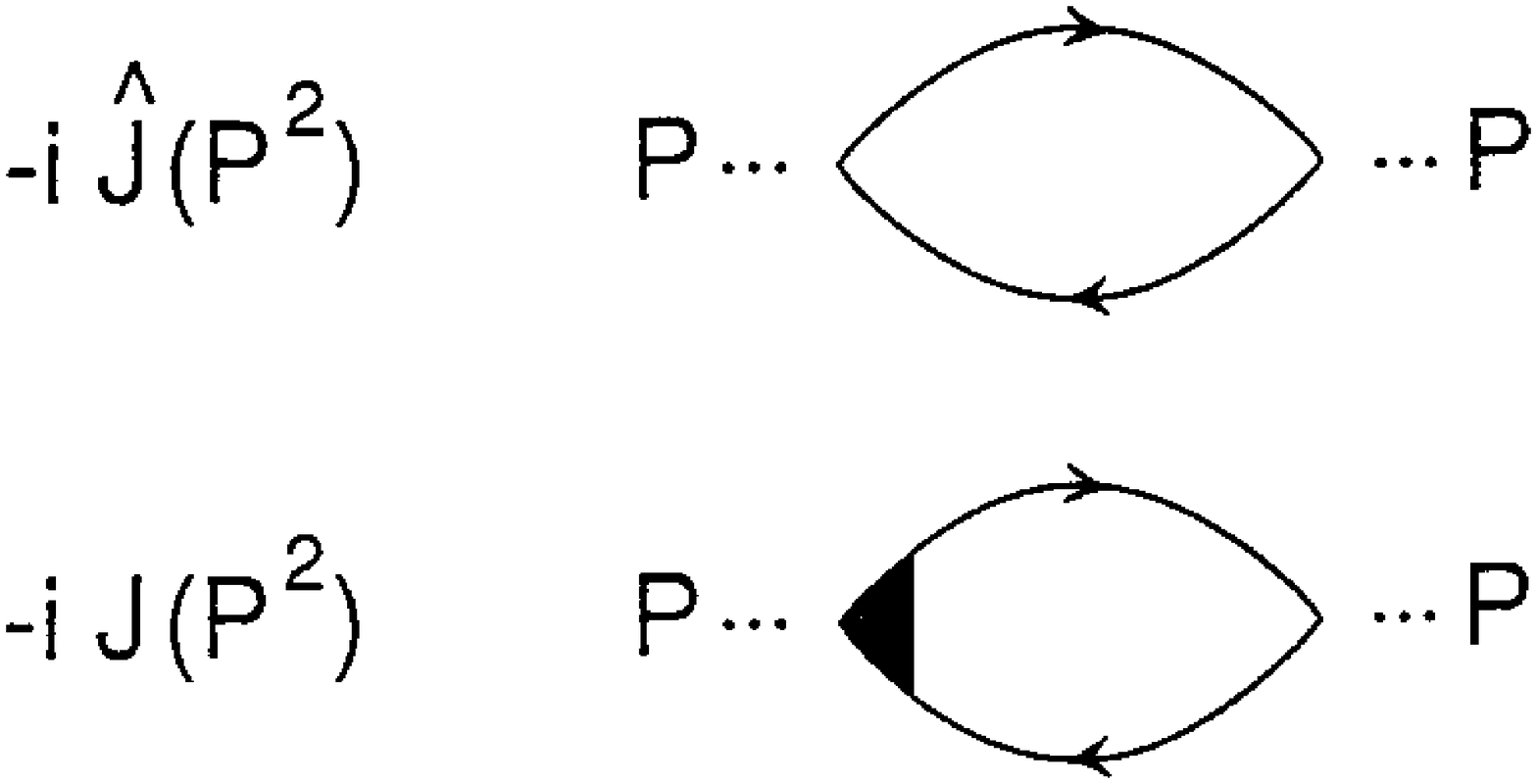}%
\caption{(a) The basic vacuum polarization diagram of the NJL
model is shown. The lines represent a constituent quark and
antiquark of mass $m(\rho_q)$.\\(b) The filled region represents a
confinement vertex used for calculations made for $\rho\leq\rho_c$
[9].}
\end{figure}

In forming the imaginary parts of the vacuum polarization
functions, the quark and antiquark go on mass shell, so that we
have $\vec k^2=(P^0)^2/4-m^2(\rho)$. It is easy to see that there
is a minimum value of $|\vec k|=p_F$, where $p_F$ is the Fermi
momentum of either the up or down ideal quark gases. For smaller
values of $|\vec k|$ the excitation is ``blocked" by the Pauli
Principle. Thus, we see that in calculating the imaginary part of
a vacuum polarization function, one obtains nonzero values above a
minimum value of $P^0$, $(P^0)_{min}^2=4(p_F^2+m^2(\rho))$, where
$p_F$ is related to the quark density by the expression
$p_F^3=(3\pi^2/2N_c)\rho_q$, given earlier.

We now consider the imaginary part of the vacuum polarization
function corresponding to scalar currents. We find in the case an
up (or down) quark is excited \be\mbox{Im}
J^S(P^2)=N_c\frac{P_0^2}{4\pi}\left(1-\frac{4m_u^2(\rho)}{P_0^2}\right)^{3/2}e^{-\vec
k^2/\alpha^2}\theta[P_0-4(p_F^2+m_u^2(\rho))]\,.\ee This
expression contains a factor of 2 arising from the flavor trace.
Here $\vec k^2=P_0^2/4-m_u^2(\rho)$ appears in the Gaussian
regulator. (We have used Gaussian regulators for most of our
calculations made using the NJL model. The value of $\alpha=0.605$
GeV yields results that are similar to those obtained with the
sharp three-momentum cutoff parameter $\Lambda=0.631$ GeV.) For
pseudoscalar mesons, we may use Eq. (2.2) with the phase-factor
exponent of 3/2 replace by 1/2 when the constituent mass is small
[10]. The real parts of the vacuum polarization functions are
obtained by means of a dispersion relation.

We now consider the calculation of density-dependent hadronic
current correlation functions. The general form of the correlator
is a transform of a time-ordered product of currents, \be C(P^2,
\rho)=i\int d^4xe^{\,ip\cdot x}<<\mbox T(j(x)j(0)>>\,,\ee where
the double bracket is a reminder that we are considering the
finite density case.

For the study of pseudoscalar states, we may consider currents of
the form $j_{P,\,i}(x)=\bar q(x)i\gamma_5\lambda^iq(x)$ where, in
the case of the $\pi$ mesons, $i=1,2,$ and 3. For the study of
pseudoscalar-isoscalar mesons, we again introduce
$j_{P,\,i}(x)=\bar q(x)\lambda^iq(x)$, but here $i=0$ for the
flavor-singlet current and $i=8$ for the flavor-octet current.

In the case of the $\pi$ mesons, the correlator may be expressed
in terms of the basic vacuum polarization function of the NJL
model, $J_P(P^2, \rho)$. Thus, \be C_\pi(P^2,\rho)=J_P(P^2,
\rho)\frac1{1-G_\pi(\rho)J_P(P^2, \rho)}\,,\ee where $G_\pi(\rho)$
is the coupling constant appropriate for our study of the $\pi$
mesons. (We have found $G_\pi(0)=13.49$\gev{-2} by fitting the
pion mass in a calculation made at $\rho=0$.)

For a study of the correlators related to the $\rho$ meson, we
introduce conserved vector currents $ j_{\mu,\,i}(x)=\bar
q(x)\gamma_\mu\lambda_iq(x)$ with $i=1,2$ and 3. In this case we
define \be
J_\rho^{\mu\nu}(P^2,\rho)=\left(g\,{}^{\mu\nu}-\frac{P\,{}^\mu
P\,{}^\nu}{P^2}\right)J_\rho(P^2,\rho)\,,\ee taking into account
the fact that the current $j_{\mu,\,i}(x)$ is conserved. We may
then use the fact that \be
J_\rho(P^2,\rho)&=&\frac13g_{\mu\nu}J_\rho^{\mu\nu}(P^2,\rho)\\
&=&\frac{2N_c}3\left[\frac{P_0^2+2m_u^2(\rho)}{4\pi}\right]
\left(1-\frac{4m_u^2(\rho)}{P_0^2}\right)^{1/2}e^{-\vec
k\,{}^2/\alpha^2}\theta[P_0^2-4(p_F^2+m_u^2(\rho))]\\&\simeq&\frac23J_\pi(P^2,\rho)\ee
and write the approximate expression \be
\mbox{Im}J_\rho(P^2,\rho)\simeq\frac23\frac{N_cP_0^2}{4\pi}\left(1-\frac{4m_u^2(\rho)}{P_0^2}\right)^{1/2}e^{-\vec
k\,{}^2/\alpha^2}\theta[P_0^2-4(p_F^2+m_u^2(\rho))]\ee for the
vacuum polarization function of the vector-isovector currents.
Here $\vec k^2=P_0^2/4-m_u^2(\rho)$ appears in the Gaussian
regulator. Thus, we define \be
C_\rho(P^2)=J_\rho(P^2)\left[\frac1{1-G_V(\rho)J_\rho(P^2)}\right]\,,\ee
where we have suppressed reference to the density dependence of
the correlator and the vacuum polarization function. We have used
$G_V(\rho)=G_V[1-0.17\rho/\rho_c]$ with $G_V=11.46$ \gev{-2} for
the calculators reported here.

The calculation of the correlator for scalar-isoscalar states is
more complex, since there are both flavor-singlet and flavor-octet
states to consider. We may define polarization functions for $u$,
$d$ and $s$ quarks: $J_u(P^2, \rho)$, $J_d(P^2, \rho)$ and
$J_s(P^2, \rho)$. These functions do not contain the factor of 2
arising from the flavor trace that was introduced when calculating
$\mbox{Im}J_\pi(P^2, \rho)$, $\mbox{Im}J^S(P^2, \rho)$ and
$\mbox{Im}J_\rho(P^2, \rho)$ earlier in this section.

In terms of these polarization functions we may then define \be
J_{00}(p^2, \rho)=\frac23[J_u(p^2, \rho)+J_d(p^2, \rho)+J_s(p^2,
\rho)]\,,\ee \be J_{08}(p^2, \rho)=\frac{\sqrt2}3[J_u(p^2,
\rho)+J_d(p^2, \rho)-2J_s(p^2, \rho)]\,,\ee and \be J_{88}(p^2,
\rho)=\frac13[J_u(p^2, \rho)+J_d(p^2, \rho)+4J_s(p^2, \rho)]\,.\ee
We also introduce the matrices \be J(p^2,
\rho)=\left[\begin{array}{cc}J_{00}(p^2, \rho)&J_{08}(p^2,
\rho)\\J_{80}(p^2, \rho)&J_{88}(p^2, \rho)\end{array}\right]\,,\ee
\be
G(\rho)=\left[\begin{array}{cc}G_{00}(\rho)&G_{08}(\rho)\\G_{80}(\rho)&G_{88}
(\rho)\end{array}\right]\,,\ee \be C(p^2,
\rho)=\left[\begin{array}{cc}C_{00}(p^2, \rho)&C_{08}(p^2,
\rho)\\C_{80}(p^2, \rho)&C_{88}(p^2, \rho)\end{array}\right]\,,\ee
and write the matrix relation \be C(p^2, \rho)=J(p^2,
\rho)[1-G(\rho)J(p^2, \rho)]^{-1}\,.\ee

\section{results of numerical calculations of pseudoscalar and vector hadronic current
correlation functions, $C_\pi(P^2)$ and $C_\rho(P^2)$}

In Fig. 2 we present values calculated for $\mbox{Im}C_\pi(P^2)$
for $\rho/\rho_c=1.2$, 2.0, 3.0, 4.0 and 5.88. The threshold for
each curve is given by $(P^0)_{min}^2=4(p_F^2+m_u^2(\rho_q))$. If
$\rho_c=2.25\rho_{NM}$, we have \be
p_F=\left(\frac{2.25}{N_c}\frac{\rho}{\rho_c}\right)^{1/3}k_F\,,\ee
where $k_F$ is the Fermi momentum of nuclear matter, $k_F=0.286$
GeV. Thus, when $\rho/\rho_c=5.88$, we find $p_F=0.439$ GeV, which
is less than the standard three-dimensional (sharp) cutoff of
$\Lambda=0.631$ GeV that is often used for the NJL model. It is
seen from the figure that there are significant nonperturbative
effects, except at $\rho/\rho_c=5.88$, where
$\mbox{Im}C_\pi(P^2)=\mbox{Im}J_\pi(P^2)$. The last result
follows, since $G_\pi(\rho)=0$ for $\rho/\rho_c=5.88$.

\begin{figure}
\includegraphics[bb=0 0 300 210, angle=0, scale=1.2]{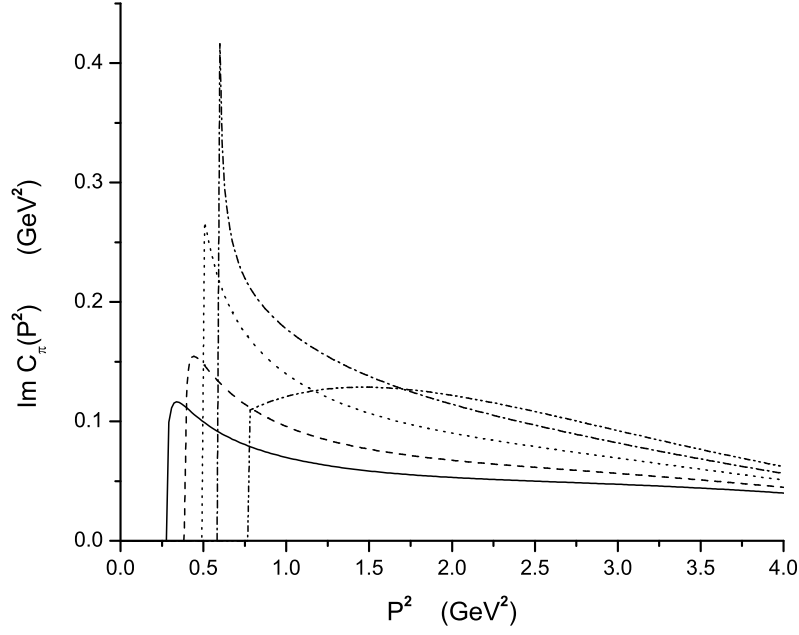}%
\caption{The figure presents values of $\mbox{Im}C_\pi(P^2)$ for
various values of $\rho/\rho_c$. Here, $\rho/\rho_c=1.2$ [solid
line], 2.0 [dashed line], 3.0 [dotted line], 4.0 [dashed-dotted
line] and 5.88 [dashed-(double)dotted line]. We have used
$G_\pi=13.51$\gev{-2}.}
\end{figure}

In Fig. 3 we show the results of a similar calculation of
$\mbox{Im}C_\rho(P^2)$. Here the thresholds for the various curves
are the same as those seen in Fig. 2.

\begin{figure}
\includegraphics[bb=0 0 300 210, angle=0, scale=1.2]{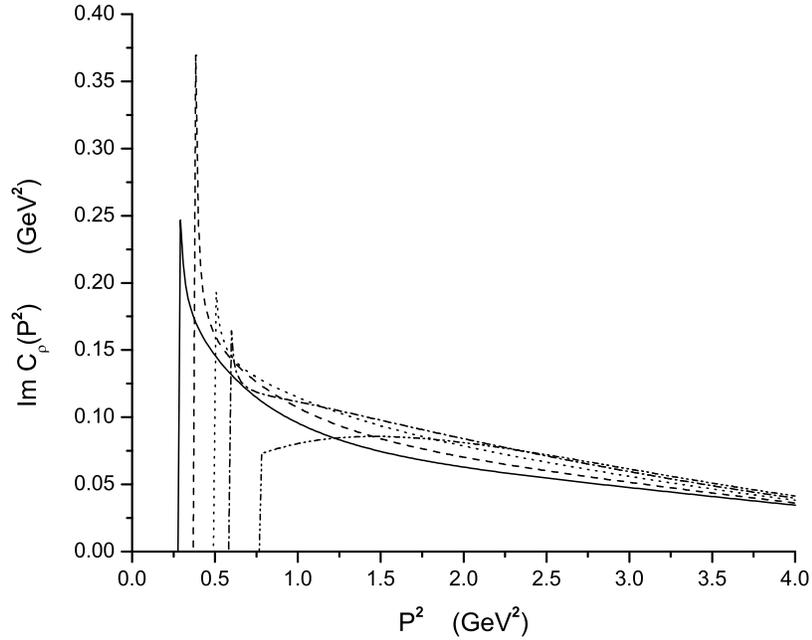}%
\caption{The figure shown the values of $\mbox{Im}C_\rho(P^2)$.
[See the caption of Fig. 2.] Here we have used
$G_V=11.46$\gev{-2}.}
\end{figure}

\section{results of numerical calculation of scalar and pseudo-scalar
correlators $C_{f_088}(P^2)$ and $C_{\eta88}(P^2)$}

In the case of the correlator of pseudoscalar-isoscalar currents,
the components $C_{\eta00}(P^2)$,
$C_{\eta08}(P^2)=C_{\eta80}(P^2)$ and $C_{\eta88}(P^2)$ are all
important. We choose to show $\mbox{Im}C_{\eta88}(P^2)$ in Fig. 4
for the values of $\rho/\rho_c=$ 1.5, 2.0, 3.0, 4.0 and 5.88. The
resonant structures have rather small widths when compared to the
features seen in Fig. 2 and 3. It is also worth noting that each
curve has two thresholds, one corresponding to $m_u(\rho)$ and the
other corresponding to $m_s(\rho)$. Thus, we have
$(P^0)_{min}^2=4[p_F^2+m_u^2(\rho)]$ or
$(P^0)_{min}^2=4[p_F^2+m_s^2(\rho)]$.

\begin{figure}
\includegraphics[bb=0 0 300 210, angle=0, scale=1.2]{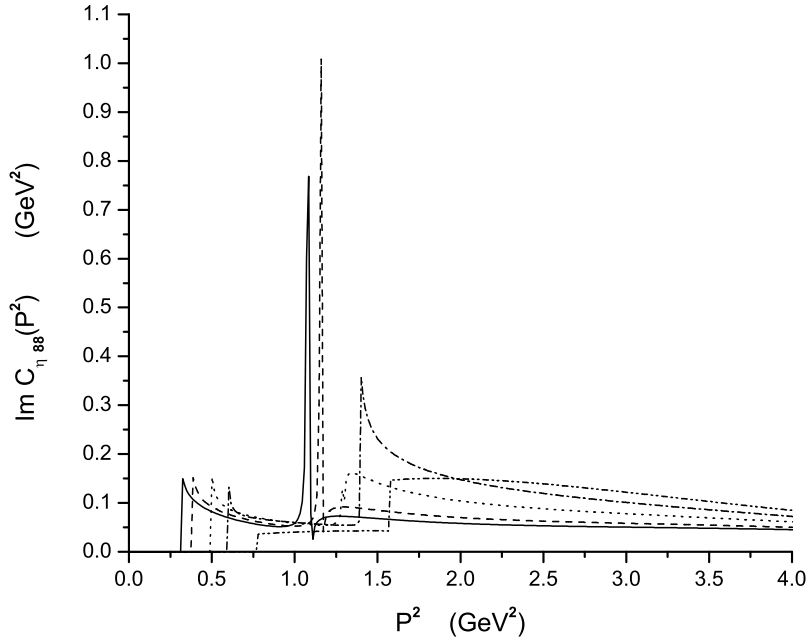}%
\caption{The figure shows the values calculated for the imaginary
part of the correlator of pseudoscalar, flavor-octet currents
$\mbox{Im}C_{\eta88}(P^2)$. Here, $\rho/\rho_c=1.5$ [solid line],
2.0 [dashed line], 3.0 [dotted line], 4.0 [dashed-dotted line],
5.0 [dashed-(double)dotted line], and 5.88 [short dashed line].}
\end{figure}

Similar remarks pertain for the scalar current correlators,
$C_{f_000}(P^2)$, $C_{f_008}(P^2)=C_{f_080}(P^2)$ and
$C_{f_088}(P^2)$. The values of $\mbox{Im}C_{f_088}(P^2)$ are
shown in Fig. 5 for various values of $\rho/\rho_c$. Note that the
resonances seen in that figure are quite narrow. That is probably
due to the different phase space behavior for the $f_0$ ($p$-wave)
and $\eta$ ($s$-wave).

\begin{figure}
\includegraphics[bb=0 0 300 210, angle=0, scale=1.2]{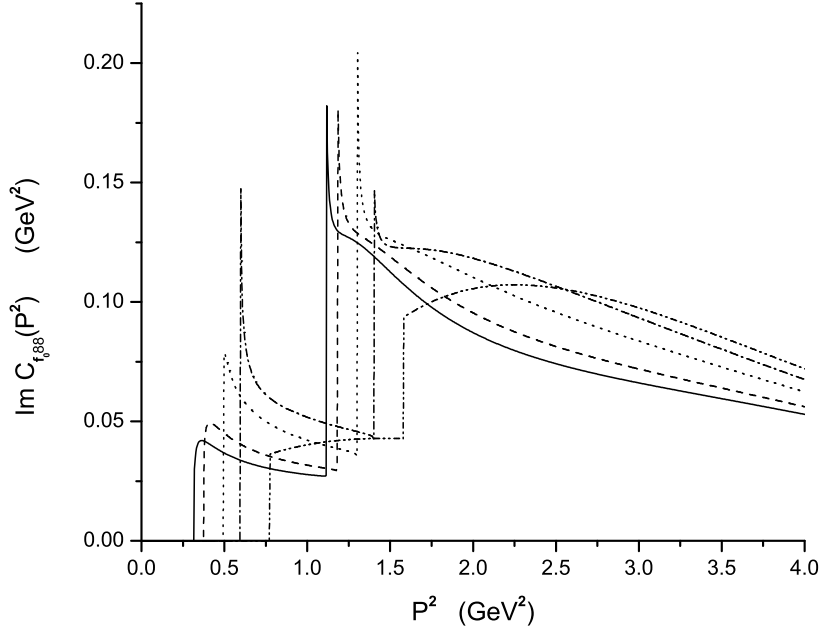}%
\caption{The values of the imaginary part of the correlator of
scalar, flavor-octet hadronic currents, $\mbox{Im}C_{f_088}(P^2)$
is shown. [See caption of Fig. 4.]}
\end{figure}

\section{results of numerical calculations with density-dependent and density-independent
NJL coupling constants}

In this section we are interested in presenting some evidence for
the density-dependent coupling constants used in our work [9]. (As
noted earlier, the argument is more easily made when we introduce
temperature-dependent coupling constants, since much more is known
concerning QCD thermodynamics at finite temperature than at finite
density.) In the case of finite density, we have assumed that the
system is weakly interacting at $\rho=5.88\rho_c$ with
$\rho_c=2.25\rho_{NM}$. (Other values for $\rho_c$ could be used,
but here we will continue to explore the consequences of the
choice made in our earlier work [9].) In Fig. 6 we compare the
results of our model for $\mbox{Im}C_\pi(P^2)$ at
$\rho/\rho_c=5.88$ with the results obtained when we use a
constant value for $G_\pi$. As seen in the figure, there is about
a factor of 3 difference in the result of the two calculations.
The difference in the two calculated results for
$\mbox{Im}C_\rho(P^2)$ seen in Fig. 7 is not quite as marked as
that seen in Fig. 6, but is still significant.

\begin{figure}
\includegraphics[bb=0 0 300 210, angle=0, scale=1.2]{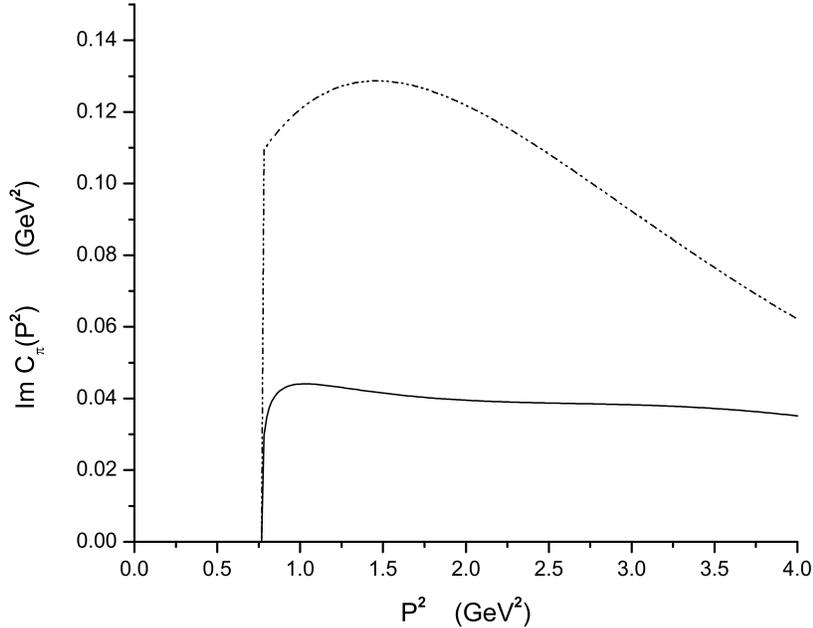}%
\caption{The solid line shows the values of
$\mbox{Im}C_{\pi}(P^2)$ for the case of density-independent NJL
coupling constants. The dashed-(double)dotted line represents
$\mbox{Im}C_{\pi}(P^2)$ calculated with the density-dependent
coupling constants $G_\pi(\rho)=G_\pi[1-0.17\rho/\rho_c]$ for
$\rho/\rho_c=5.88$.}
\end{figure}

\begin{figure}
\includegraphics[bb=0 0 300 210, angle=0, scale=1.2]{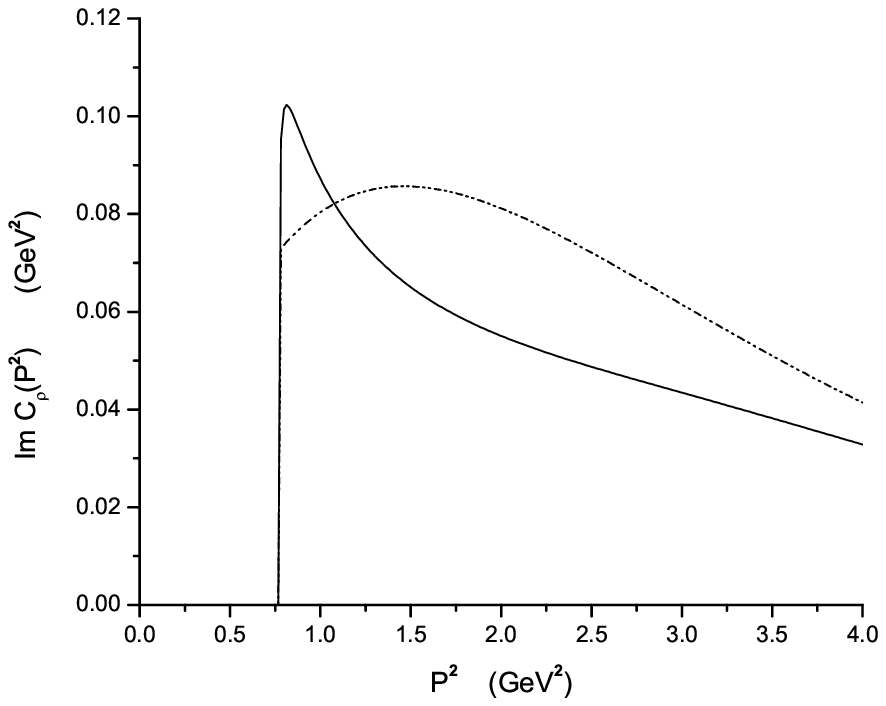}%
\caption{The solid line shows the values of
$\mbox{Im}C_{\rho}(P^2)$ for the case of density-independent NJL
coupling constants. The dashed-(double)dotted line represents
$\mbox{Im}C_{\rho}(P^2)$ calculated with the density-dependent
coupling constants $G_V(\rho)=G_V[1-0.17\rho/\rho_c]$ for
$\rho/\rho_c=5.88$.}
\end{figure}

In Fig. 8 we compare the results for $\mbox{Im}C_{\eta88}(P^2)$ at
$\rho/\rho_c=5.88$ for the two methods of calculation. A rather
dramatic difference is seen in Fig. 8, where we see that the
calculation with constant values of $G_{00}^P$, $G_{08}^P$ and
$G_{88}^P$ leads to a resonance at $P^2=1.53$\gev2. In Fig. 9 we
show similar results for $\mbox{Im}C_{f_088}(P^2)$ with a
resonance at $P^2=1.58$\gev2 in the case that constant values of
$G_{00}^P$, $G_{08}^P$ and $G_{88}^P$ are used.

\begin{figure}
 \includegraphics[bb=0 0 300 210, angle=0, scale=1.2]{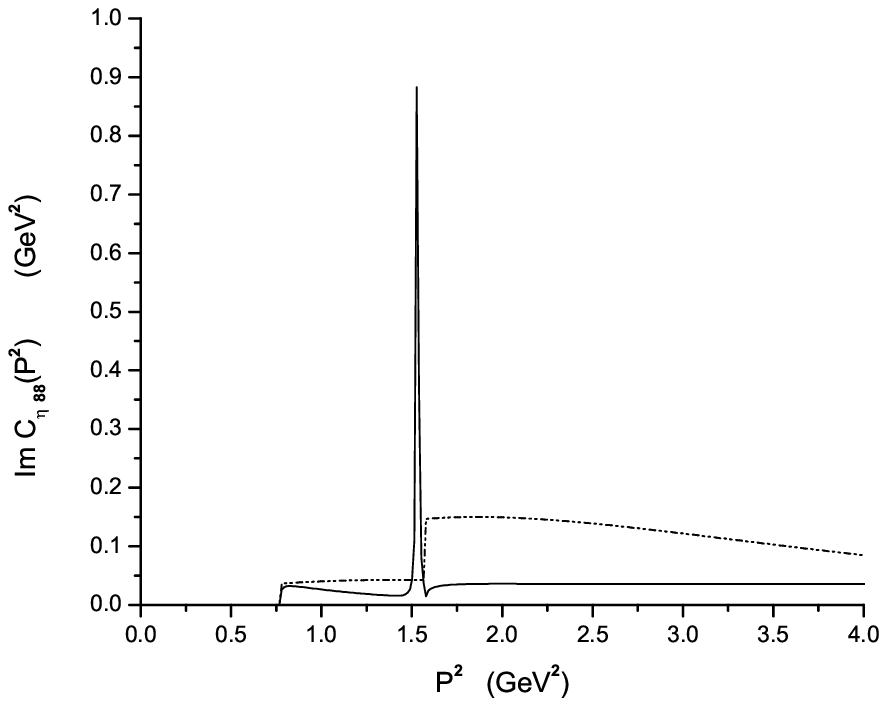}%
 \caption{The figure shows $\mbox{Im}C_{\eta88}(P^2)$ calculated for $\rho/\rho_c=5.88$.
 The dashed-(double)dotted line is the result of using our density-dependent coupling parameters,
 while the solid line is the result when density-independent coupling parameters are used.}
 \end{figure}

 \begin{figure}
 \includegraphics[bb=0 0 300 210, angle=0, scale=1.2]{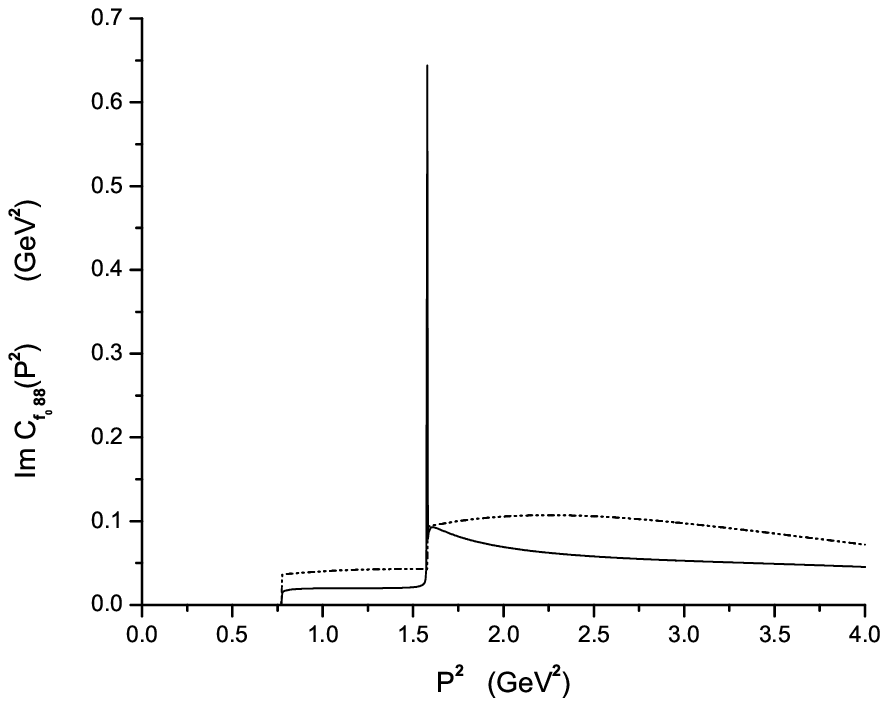}%
 \caption{The figure shows $\mbox{Im}C_{f_088}(P^2)$} calculated for
 $\rho/\rho_c=5.88$. [See the caption of Fig. 8.]
 \end{figure}

\section{discussion}

In this work we have assumed that deconfinement takes place at
$\rho_c=2.25\rho_{NM}$ and that the quark gluon plasma is weakly
interacting at $\rho=5.88\rho_c=13.2\rho_{NM}$. This particular
model was explored in Ref. [9] where we calculated the masses of
several mesons and their radial excitations for various matter
densities, $\rho\leq\rho_c$, making use of our generalized NJL
model with confinement. In the present work we have put forth some
evidence that the NJL coupling constants should be density
dependent to obtain a consistent formalism. We may consider what
results would be obtained if we used a larger value of the density
for the confinement-deconfinement transition. We have investigated
the values of $\rho_c=3.0\rho_{NM}$, $\rho_c=4.0\rho_{NM}$ and
$\rho_c=5.0\rho_{NM}$ and in each case have assumed that the
quark-gluon plasma is weakly interacting for $\rho=6.0\rho_c$. (We
should note that, if $\rho_c$ is made large, the value calculated
for $p_F$ will become larger than the value of the three-momentum
NJL cutoff, $\Lambda=0.631$ GeV, that is often used. However, our
computer code still provides results under that circumstance,
since we use a Gaussian regulator.) It is found in all the
calculations made at the values of $\rho$ for which we have
assumed the system to be weakly interacting, that the values of
the correlators calculated with constant values of the coupling
constants differ significantly from the results obtained with the
density-dependent coupling constants. We suggest that this matter
should be resolved before we undertake studies of diquark
condensation at high densities and related matters.

One characteristic of our calculation of hadronic current
correlators at finite temperature and finite density in the
deconfined phase is the appearance of complex resonance structure
in some cases. That feature is in general agreement with the
observation made in Ref. [26] ``...that the correlators possess a
nontrivial structure in the deconfined phase." Meson correlators
were studied in finite temperature lattice QCD in Ref. [27], in
which the authors state, ``...Below $T_c$ we observe little change
in the meson properties as compared with $T=0$. Above $T_c$ we
observe new features: chiral symmetry restoration and signals of
plasma formation, but also an indication of persisting ``mesonic"
(metastable) states and different temporal and spacial masses in
the mesonic channels. This suggests a complex picture of QGP in
the region (1-1.5)$T_c$".




\end{document}